\documentclass[12pt]{article}
\usepackage{epsfig}

\def\beq{\begin{equation}}
\def\eeq#1{\label{#1}\end{equation}}
\def\eeqn{\end{equation}}

\def\beqa{\begin{eqnarray}}
\def\eeqa#1{\label{#1}\end{eqnarray}}
\def\eeqan{\end{eqnarray}}

\let\bar=\overbar

\def\Dslash{\not{\hbox{\kern-4pt $D$}}}
\def\dslash{\not{\hbox{\kern-2pt $\del$}}}

\def\msb{{\bar{\ssstyle M \kern -1pt S}}}

\usepackage{fancyhdr,graphicx}
\def\Title#1{\begin{center} {\Large {\bf #1} } \end{center}}

\begin{document}

\Title{Phase structure of hadronic and Polyakov-loop extended NJL model at finite isospin density}

\bigskip\bigskip


\begin{raggedright}

{\it 
Rafael Cavagnoli${}^{\;\; 1,2}$ $\,$ Debora P. Menezes$^{\;\; 1}$ and Constan\c{c}a Provid\^encia$^{\;\; 3}$ \\
\bigskip
${}^{1}$Departamento de F\'{\i}sica, Universidade Federal de Santa Catarina,  
Florian\'opolis, CP. 476, CEP 88.040-900, Brasil \\
\bigskip
${}^{2}$Instituto de F\'isica e Matem\'atica, Universidade Federal de Pelotas, CP. 354, CEP 96001-970, Pelotas, Brasil \\
\bigskip
${}^{3}$Centro de F\'isica Computacional, Department of Physics, University of Coimbra, P3004-516
Coimbra, Portugal
}

\end{raggedright}

\section{Introduction}

It is believed that there exists a rich phase structure of quantum chromodynamics (QCD) at finite temperature and baryon density, namely, the deconfinement process from hadron gas to quark–gluon plasma, the transition from chiral symmetry breaking phase to the symmetry restoration phase, and the color superconductivity at low temperature and high baryon density. In the present work we study the hadron-quark phase transition by investigating the binodal surface and extending it to finite temperature in order to mimic the QCD phase diagram \cite{rafa-2011}. In order to obtain these conditions we use different models for the two possible phases, namely the quark and hadron phases. The phase separation boundary (binodal) is determined by the Gibbs criteria \cite{landau-gibbs-muller} for phase equilibrium. The boundaries of the mixed phase and the related critical points for symmetric and asymmetric matter are obtained. Isospin effects appear to be rather significant. The critical endpoint (CEP) \cite{cep-studied}
 and the phase structure are also studied in the Polyakov-loop extended Nambu-Jona-Lasinio model.

\section{Formalism}

The deconfinement hadron-quark phase transition can be studied within the framework of a two-phase model where the equations of state (EoS) for the hadron and quark phases are built using two different models. For the hadronic asymmetric matter we have used the relativistic non-linear Walecka model (NLWM) \cite{bb}. In this model the hadrons are coupled to a neutral scalar $\sigma$, isoscalar-vector $\omega^\mu$ and isovector-vector $\vec \rho^\mu$  meson fields. The Lagrangian density used in the hadron phase and also the EoS for the quark phase (the MIT bag model $+$ gluons) are presented in reference \cite{rafa-2011}. 

Due to limitations of these simple models, some features of the QCD phase diagram can be studied using the Nambu-Jona-Lasinio model (NJL), a quark model which describes the correct chiral properties not present in the MIT bag model. By introducing static gluonic degrees of freedom in the NJL model through an effective gluon potential in terms of Polyakov loops it is possible to study features of both chiral symmetry breaking and deconfinement in the PNJL model. For more details on the Nambu-Jona-Lasinio model coupled to the Polyakov Loops (PNJL) see reference \cite{pedro-costa-1}.



\section{Results and Conclusions}

\begin{figure}[htb]
\begin{center}
\begin{tabular}{cc}
\includegraphics[width=7.2cm]{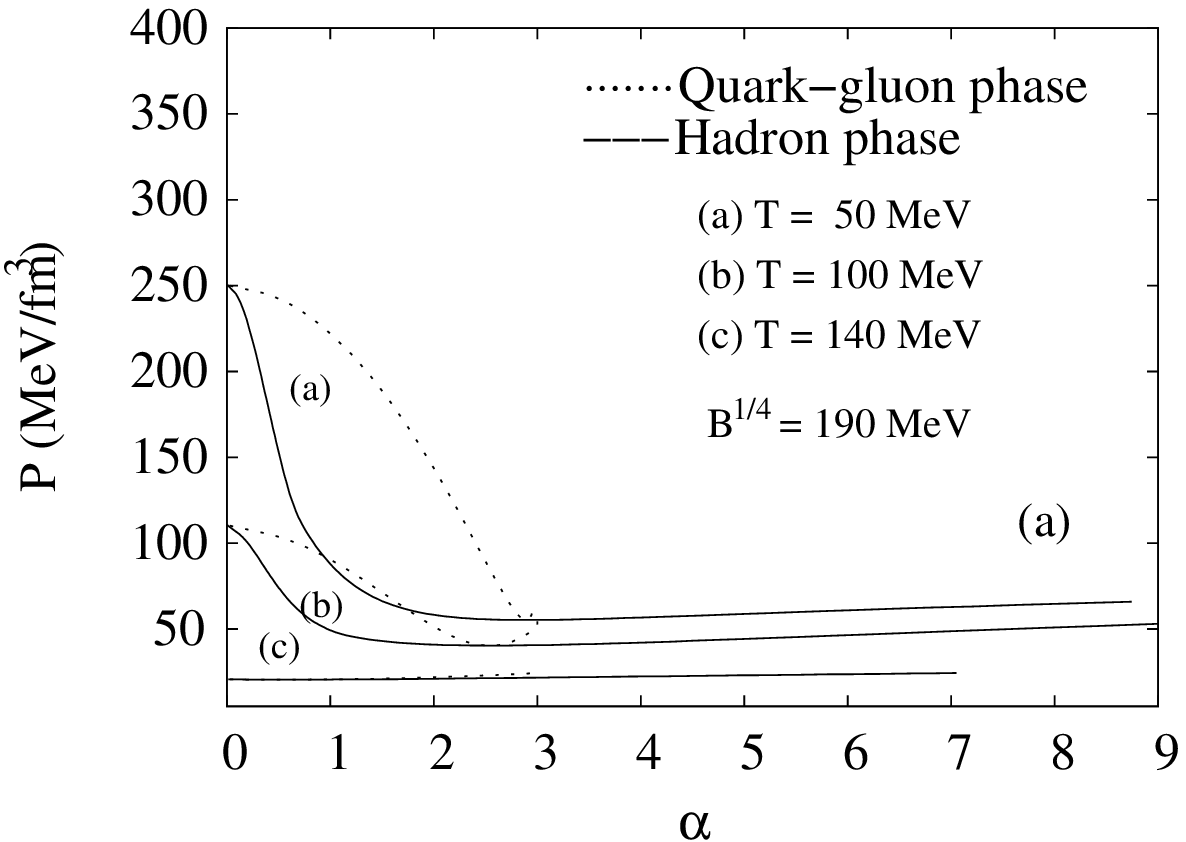} &         
\includegraphics[width=7.2cm]{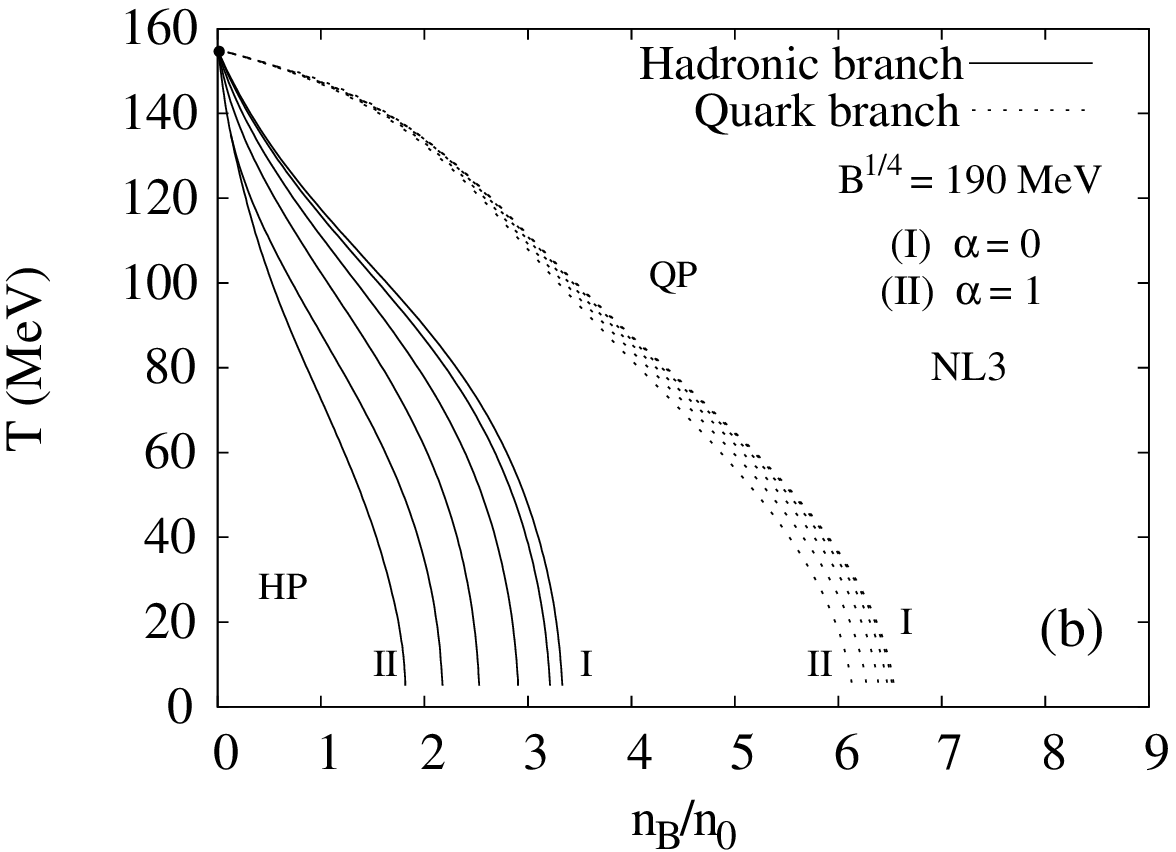}\\         
\includegraphics[width=7.2cm]{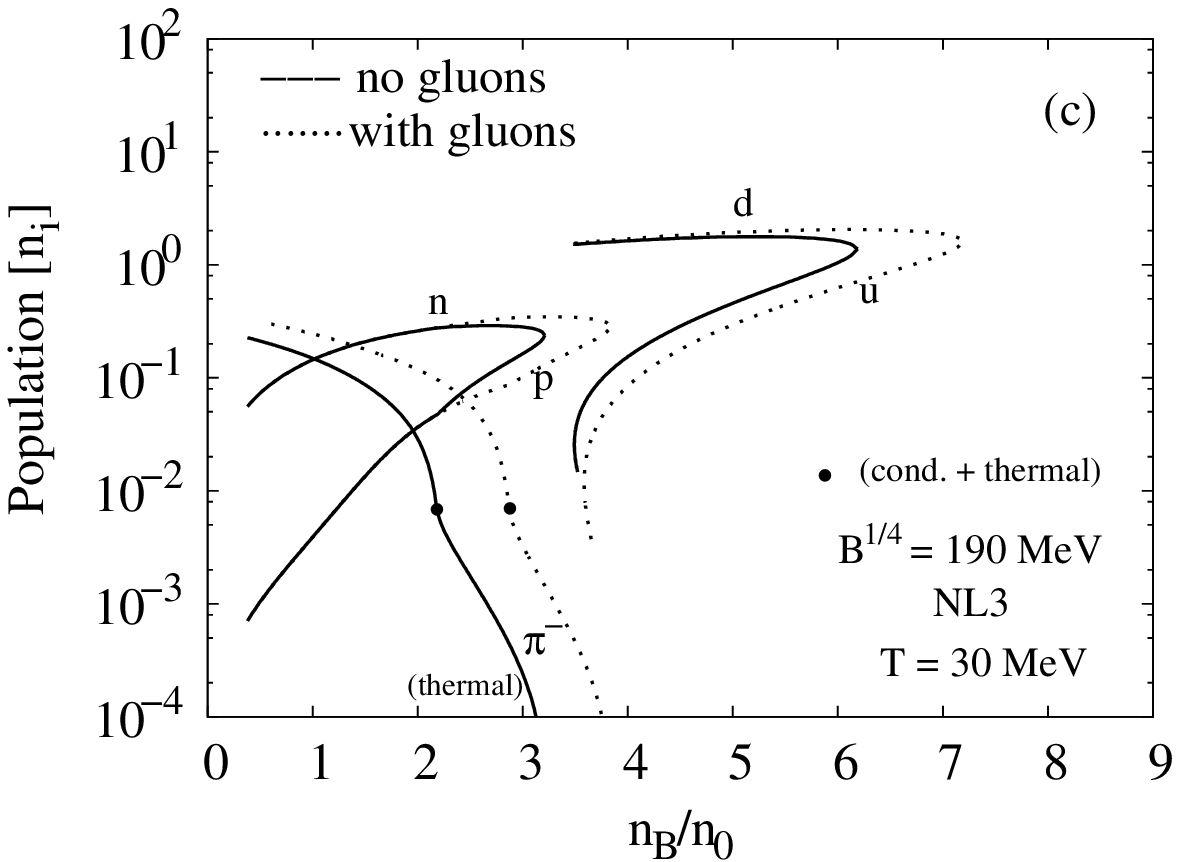} &         
\includegraphics[width=7.2cm]{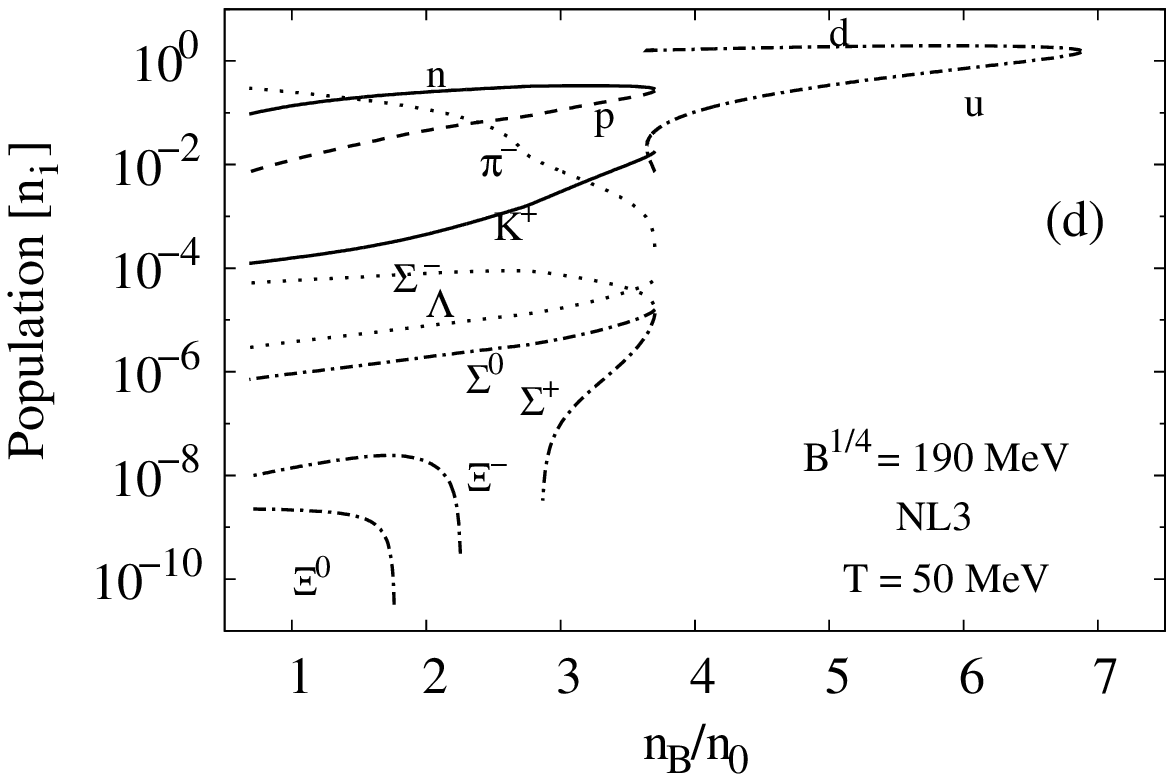}\\         
\end{tabular}
\caption{The binodal surface and the particle population in the two-phase model. We have used the NL3 parameter set for the Walecka model and for the MIT model we have used $B^{1/4} = 190$~MeV for the bag constant. Hyperons are also included with zero total strangeness.}
\label{figs1}
\end{center}
\end{figure}

\begin{figure}[htb]
\begin{center}
 
\begin{tabular}{cc}
\includegraphics[width=7.2cm]{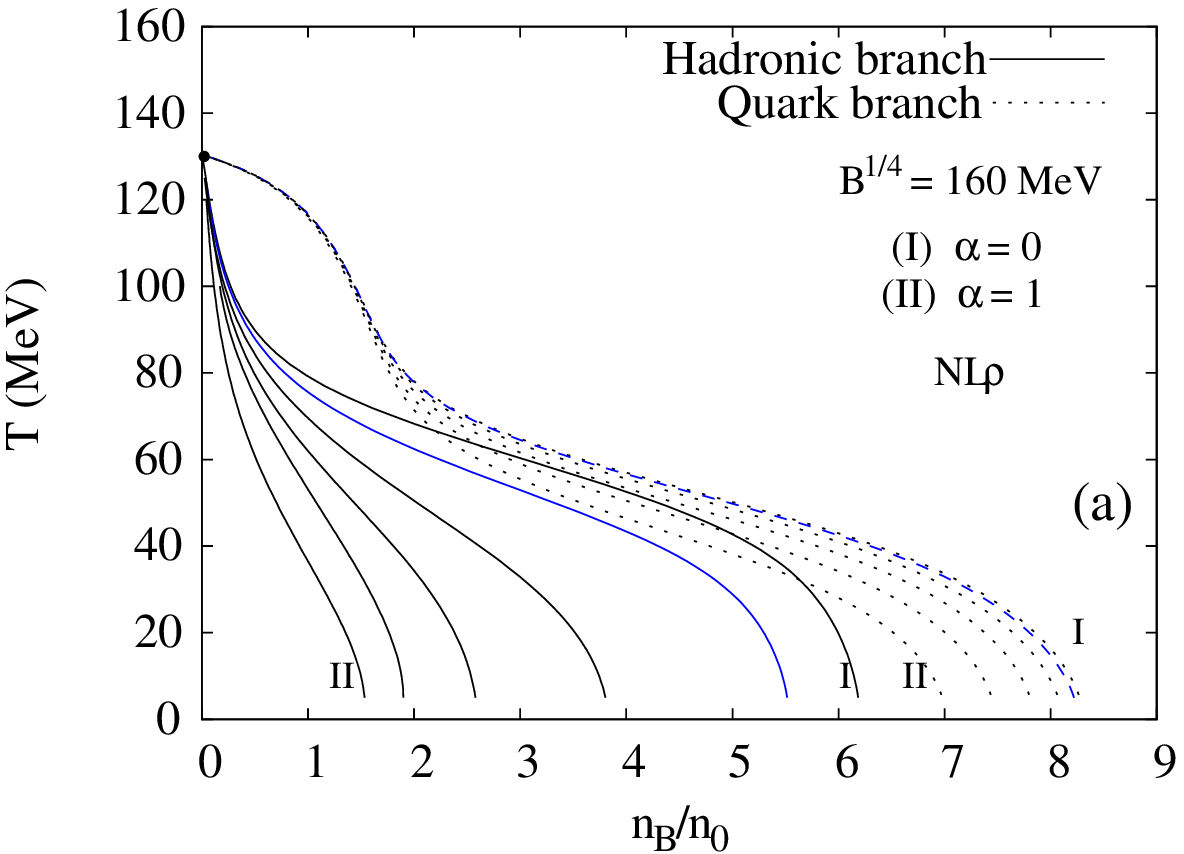} &         
\includegraphics[width=7.2cm]{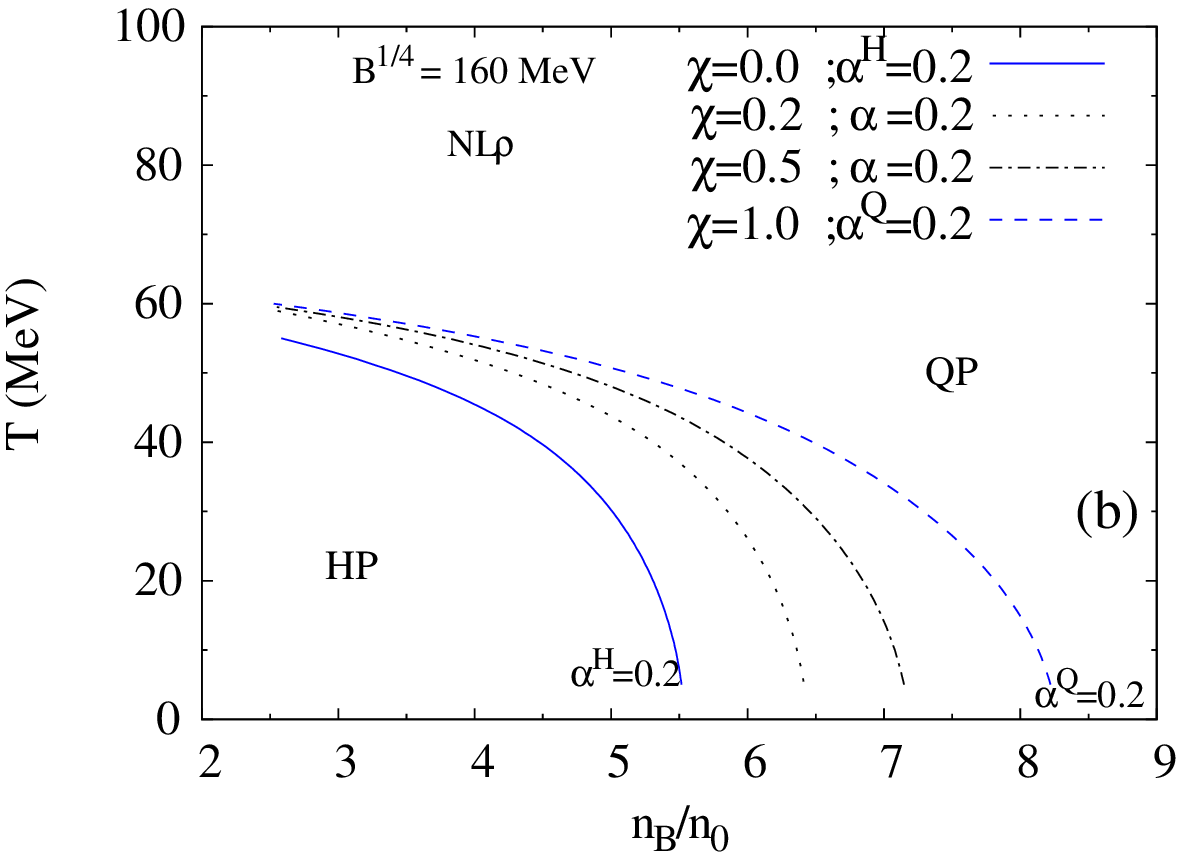}\\         
\end{tabular}
 \caption{Particle population and the binodal surface in the two phase model.}
 \label{figs2}
\end{center}
\end{figure}

\begin{figure}[htb]
\begin{center}
 
\begin{tabular}{cc}
\includegraphics[width=7.2cm]{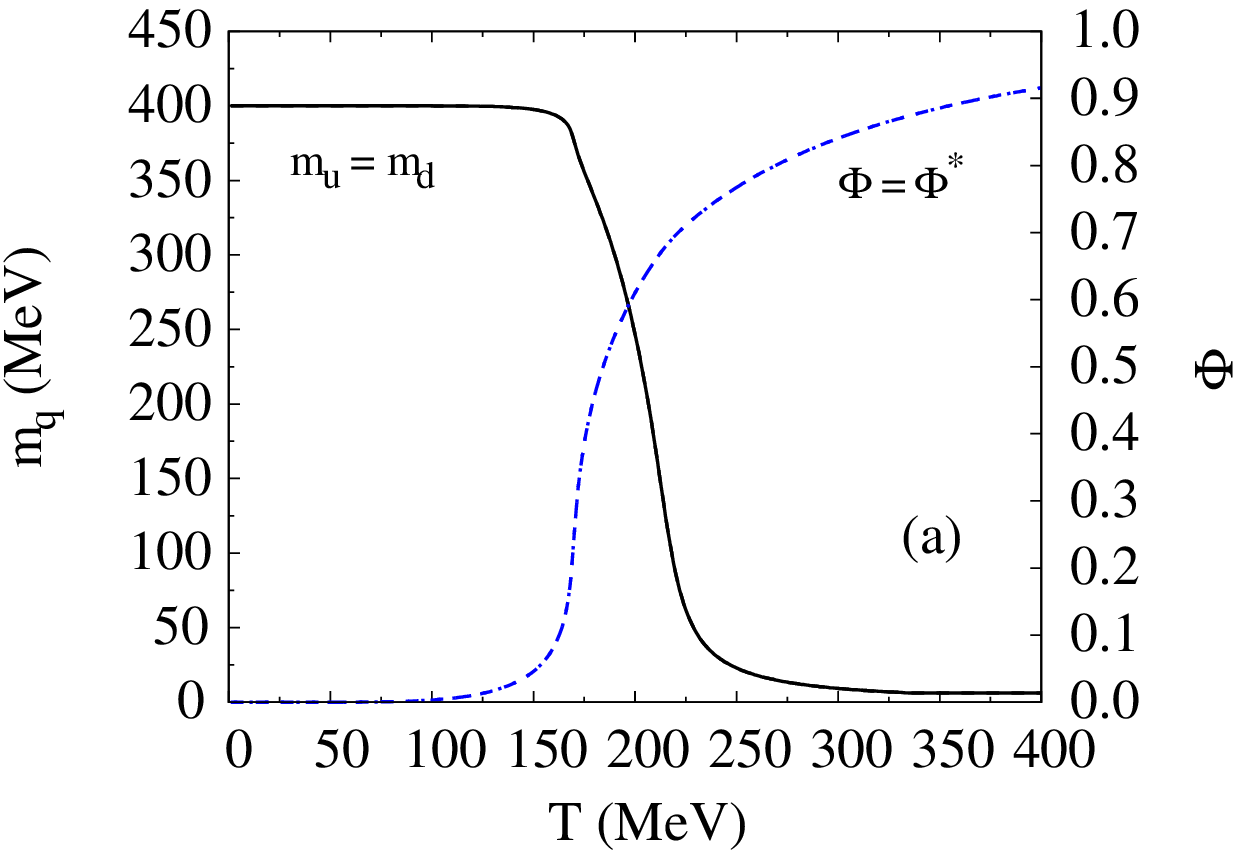} &
\includegraphics[width=7.2cm]{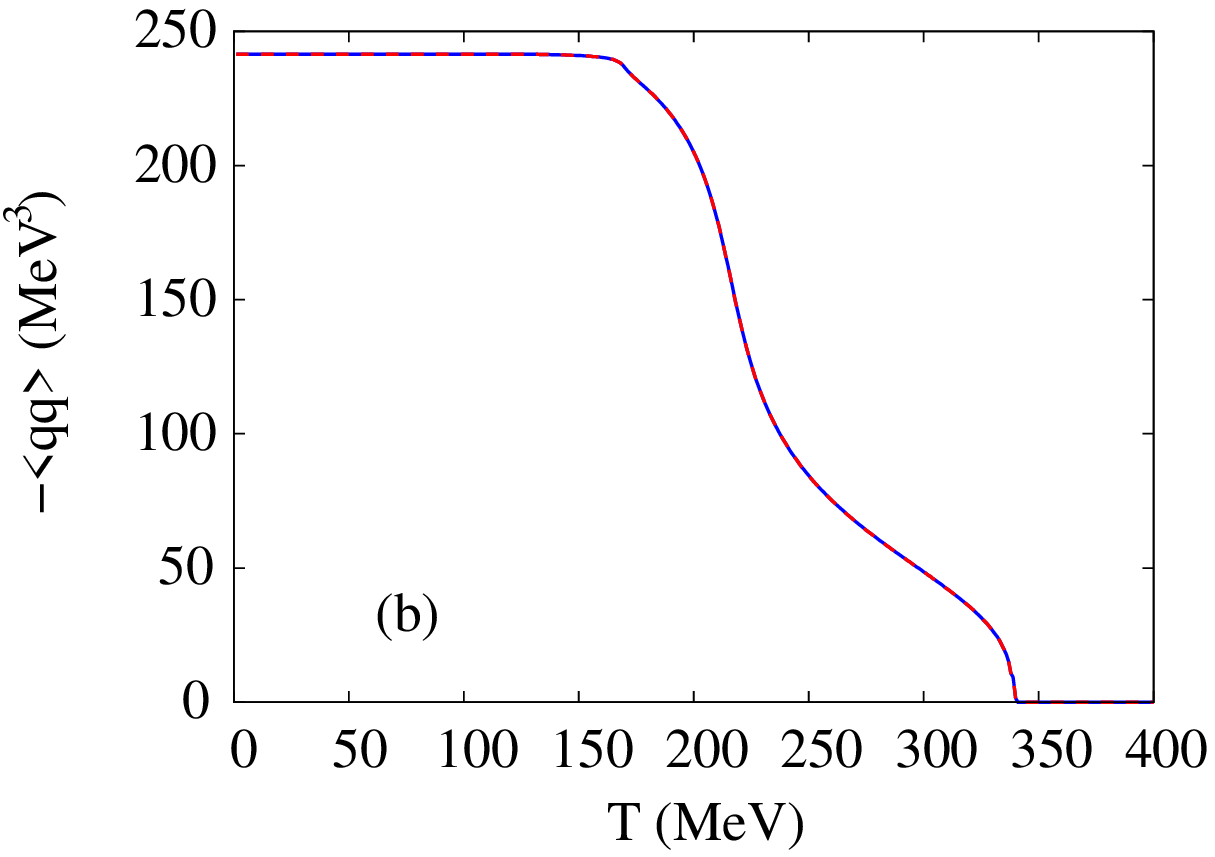}\\
\end{tabular}
 \caption{The quark masses, the Polyakov loop field $\Phi$ and the quark condensate as a function of the temperature at zero chemical potential.}
 \label{figs3}
\end{center}
\end{figure}

Figures \ref{figs1} and \ref{figs2} show the two-phase model results while the preliminary results for the PNJL model are shown in figure \ref{figs3}, for the two flavor case.

With a very simple quark model (MIT) one can obtain interesting results as seen from the figures \ref{figs1} and \ref{figs2} in its simplest form - the $SU(2)$ sector. In fig. \ref{figs1}~(a) one sees the binodal surface projected onto the \{$\alpha, P$\} plane for different temperatures, where $\alpha$ and $P$ are the asymmetry parameter and the pressure, where the asymmetry parameter takes into account the isospin of the system. It is possible to see the change of binodal shape with temperature. The enclosed area becomes smaller when temperature is increased until $T_{c}$, where the two lines combine to a single line at constant pressure. For the bag parameter between $B^{1/4} = 190 - 210$~MeV, $T_{c}$ lies between $150-170$~MeV. Figure \ref{figs1}~(b) shows the projection of several binodal branches at different $\alpha$ onto the \{$n_B, T$\} plane. At zero baryon density (zero chemical potential) the hadron-quark phase transition temperature lies between $T = 150 - 170$~MeV, within the expected range. 
Figure \ref{figs1}~(c) shows the particle population for a simple case, showing the onset of $\pi^-$ condensation. The particle population including the hyperons from the baryon octet is shown in fig. \ref{figs1}~(d), considering zero net strangeness in both phases. 

Using the NL$\rho$ parameter set for the hadron EoS one sees from figs. \ref{figs2}~(a) and \ref{figs2}~(b) the indication of an interesting region where the phase transition probably occurs and can be probed by intermediate energy heavy-ion collisions. This region is located in the range $n_B = 2 - 4 n_0$ and $T = 50 - 65$~MeV where some calculations indicate the existence of the critical point. 

\indent Some of the preliminary results using the PNJL model for the $SU(2)$ sector are presented in figures \ref{figs3} as the quark masses, the Polyakov loop field $\Phi$ and the quark condensates as a function of the temperature at zero chemical potential. The chiral phase transition takes place at 170 MeV for a given parametrization according to fig. \ref{figs3}~(a). From figures \ref{figs3}~(a) and (b) it is possible to see that the Polyakov loop field and the quark condensate vary continuously with temperature at $\mu = 0$ and there exist sharp increase (decrease) at high temperatures. A similar behavior is observed in the pressure and entropy curves, a sharp increase near the transition temperature saturating at the corresponding ideal gas limit at higher temperature.

\section{Acknowledgments}

We are grateful to Pedro Costa for providing us part of the Fortran code.

\end{document}